\documentclass{epl}
\usepackage{graphicx}

\title{Temporal evolution of the ``thermal'' and ``superthermal''
  income classes in the USA during 1983--2001}
  \shorttitle{``Thermal'' and ``superthermal'' income classes}

\author{A. Christian Silva \and Victor M. Yakovenko}
\shortauthor{A. C. Silva \etal}

\institute{Department of Physics, University of Maryland, College
  Park, MD 20742-4111, USA}


\date{31 October 2004}

\begin{document}

\maketitle

\begin{abstract}
  Personal income distribution in the USA has a well-defined two-class
  structure.  The majority of population (97--99\%) belongs to the
  lower class characterized by the exponential Boltzmann-Gibbs
  (``thermal'') distribution, whereas the upper class (1--3\% of
  population) has a Pareto power-law (``superthermal'') distribution.
  By analyzing income data for 1983--2001, we show that the
  ``thermal'' part is stationary in time, save for a gradual increase
  of the effective temperature, whereas the ``superthermal'' tail
  swells and shrinks following the stock market.  We discuss the
  concept of equilibrium inequality in a society, based on the
  principle of maximal entropy, and quantitatively show that it
  applies to the majority of population.
\end{abstract}

Attempts to apply the methods of exact sciences, such as physics, to
describe a society have a long history \cite{Ball}.  At the end of the
19th century, Italian physicist, engineer, economist, and sociologist
Vilfredo Pareto suggested that income distribution in a society is
described by a power law \cite{Pareto}.  Modern data indeed confirm
that the upper tail of income distribution follows the Pareto law
\cite{Champernowne,Aoki,Souma,Gallegati,Australia}.  However, the
majority of population does not belong there, so characterization and
understanding of their income distribution remains an open problem.
Dr\u{a}gulescu and Yakovenko \cite{Yakovenko-money} proposed that the
equilibrium distribution should follow an exponential law analogous to
the Boltzmann-Gibbs distribution of energy in statistical physics.
The first factual evidence for the exponential distribution of income
was found in Ref.\ \cite{Yakovenko-income}.  Coexistence of the
exponential and power-law parts of the distribution was recognized in
Ref.\ \cite{Yakovenko-wealth}.  However, these papers, as well as
Ref.\ \cite{Yakovenko-survey}, studied the data only for a particular
year.  Here we analyze temporal evolution of the personal income
distribution in the USA during 1983--2001.  We show that the US
society has a well-defined two-class structure.  The majority of
population (97--99\%) belongs to the lower class and has a very stable
in time exponential (``thermal'') distribution of income.  The upper
class (1--3\% of population) has a power-law (``superthermal'')
distribution, whose parameters significantly change in time with the
rise and fall of stock market. Using the principle of maximal entropy,
we discuss the concept of equilibrium inequality in a society and
quantitatively show that it applies to the bulk of population.  Most
of academic and government literature on income distribution and
inequality \cite{Kakwani,Cowell,Atkinson,Petska} does not attempt to
fit the data by a simple formula.  When fits are performed, usually
the log-normal distribution \cite{Gibrat} is used for the lower part
of the distribution \cite{Souma,Gallegati,Australia}.  Only recently
the exponential distribution started to be recognized in income
studies \cite{Nirei,Mimkes}, and models showing formation of two
classes started to appear \cite{West,Wright}.

Let us introduce the probability density $P(r)$, which gives the
probability $P(r)\,dr$ to have income in the interval $(r,r+dr)$.  The
cumulative probability $C(r)=\int_r^\infty dr'P(r')$ is the
probability to have income above $r$, $C(0)=1$.  By analogy with the
Boltzmann-Gibbs distribution in statistical physics
\cite{Yakovenko-money,Yakovenko-income}, we consider an exponential
function $P(r)\propto\exp(-r/T)$, where $T$ is a parameter analogous
to temperature.  It is equal to the average income $T=\langle
r\rangle=\int_0^\infty dr'r'P(r')$, and we call it the ``income
temperature.''  When $P(r)$ is exponential, $C(r)\propto\exp(-r/T)$ is
also exponential.  Similarly, for the Pareto power law $P(r)\propto
1/r^{\alpha+1}$, $C(r)\propto 1/r^\alpha$ is also a power law.

\begin{figure}
\twofigures[width=0.37\linewidth,angle=-90]{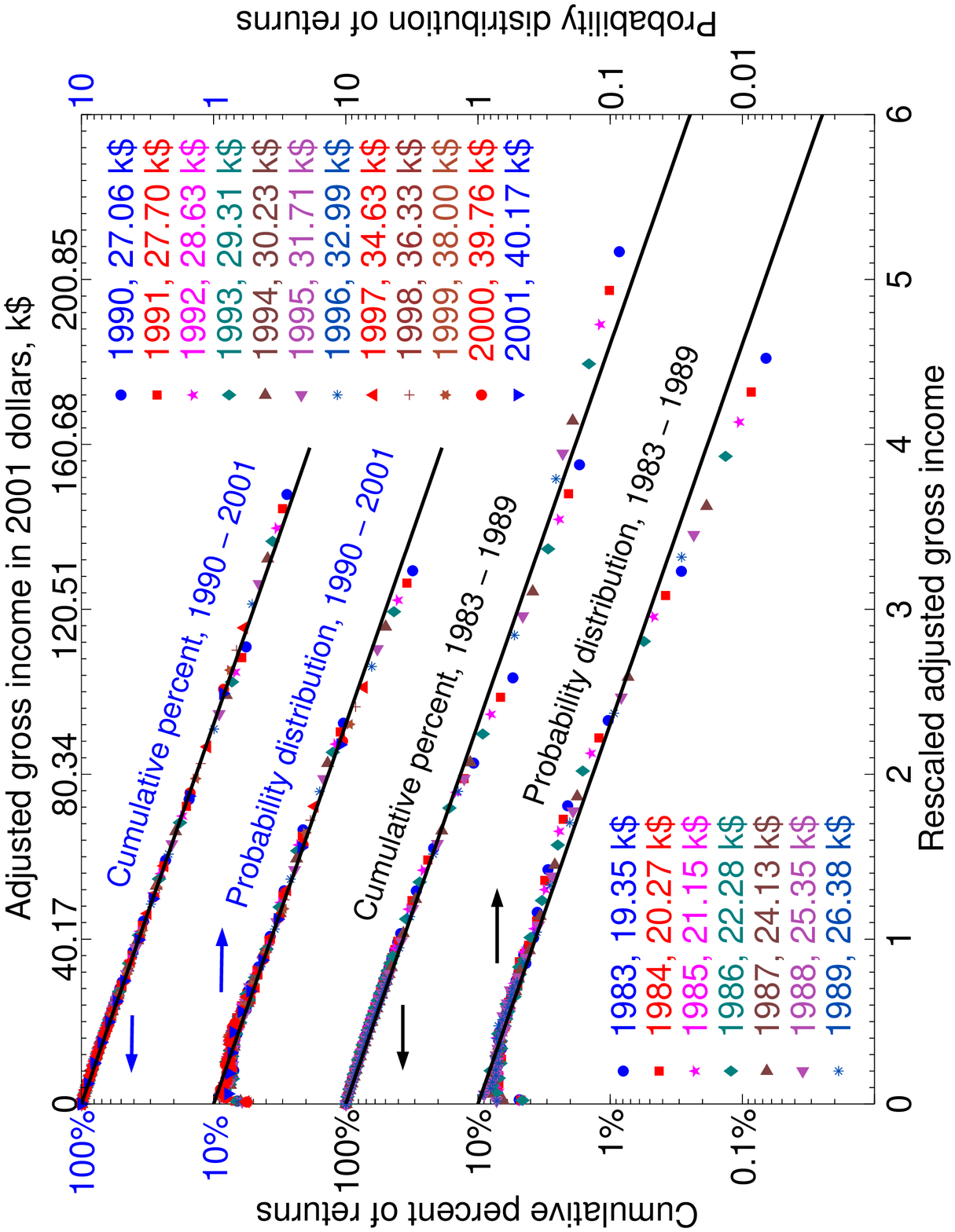}{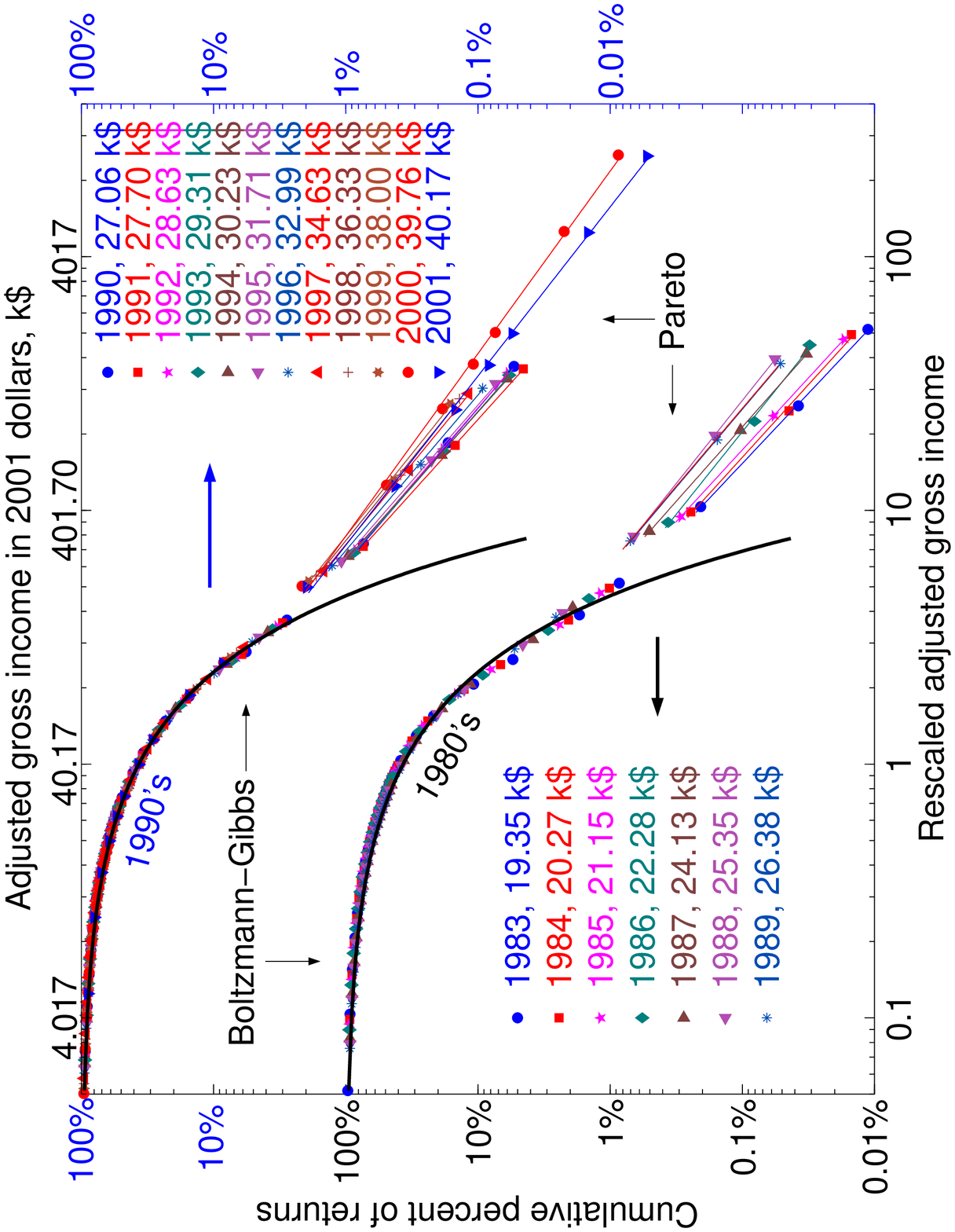}
\caption{
  Cumulative probability $C(r)$ and probability density $P(r)$ plotted
  in the log-linear scale vs.\ $r/T$, the annual personal income $r$
  normalized by the average income $T$ in the exponential part of the
  distribution. The IRS data points are for 1983--2001, and the
  columns of numbers give the values of $T$ for the corresponding
  years.}
\label{fig:LogLin}
\caption{
  Log-log plots of the cumulative probability $C(r)$ vs.\ $r/T$ for a
  wider range of income $r$.}
\label{fig:LogLog}
\end{figure}

We analyze the data \cite{IRS} on personal income distribution
compiled by the Internal Revenue Service (IRS) from the tax returns in
the USA for the period 1983--2001 (presently the latest available
year).  The publicly available data are already preprocessed by the
IRS into bins and effectively give the cumulative distribution
function $C(r)$ for certain values of $r$.  First we make the plots of
$\log C(r)$ vs.\ $r$ (the log-linear plots) for each year.  We find
that the plots are straight lines for the lower 97--98\% of
population, thus confirming the exponential law.  From the slopes of
these straight lines, we determine the income temperatures $T$ for
each year.  In Fig.\ \ref{fig:LogLin}, we plot $C(r)$ and $P(r)$ vs.\
$r/T$ (income normalized to temperature) in the log-linear scale.  In
these coordinates, the data sets for different years collapse onto a
single straight line.  (In Fig.\ \ref{fig:LogLin}, the data lines for
1980s and 1990s are shown separately and offset vertically.)  The
columns of numbers in Fig.\ \ref{fig:LogLin} list the values of the
annual income temperature $T$ for the corresponding years, which
changes from 19 k\$ in 1983 to 40 k\$ in 2001.  The upper horizontal
axis in Fig.\ \ref{fig:LogLin} shows income $r$ in k\$ for 2001.

In Fig.\ \ref{fig:LogLog}, we show the same data in the log-log scale
for a wider range of income $r$, up to about $300T$.  Again we observe
that the sets of points for different years collapse onto a single
exponential curve for the lower part of the distribution, when plotted
vs.\ $r/T$.  However, above a certain income $r_*\approx4T$, the
distribution function changes to a power law, as illustrated by the
straight lines in the log-log scale of Fig.\ \ref{fig:LogLog}.  Thus
we observe that income distribution in the USA has a well-defined
two-class structure.  The lower class (the great majority of
population) is characterized by the exponential, Boltzmann-Gibbs
distribution, whereas the upper class (the top few percent of
population) has the power-law, Pareto distribution.  The intersection
point of the exponential and power-law curves determines the income
$r_*$ separating the two classes.  The collapse of data points for
different years in the lower, exponential part of the distribution in
Figs.\ \ref{fig:LogLin} and \ref{fig:LogLog} shows that this part is
very stable in time and, essentially, does not change at all for the
last 20 years, save for a gradual increase of temperature $T$ in
nominal dollars.  We conclude that the majority of population is in
statistical equilibrium, analogous to the thermal equilibrium in
physics.  On the other hand, the points in the upper, power-law part
of the distribution in Fig.\ \ref{fig:LogLog} do not collapse onto a
single line.  This part significantly changes from year to year, so it
is out of statistical equilibrium.  A similar two-part structure in
the energy distribution is often observed in physics, where the lower
part of the distribution is called ``thermal'' and the upper part
``superthermal'' \cite{superthermal}.

\begin{figure}
\twofigures[width=0.5\linewidth,clip=true]{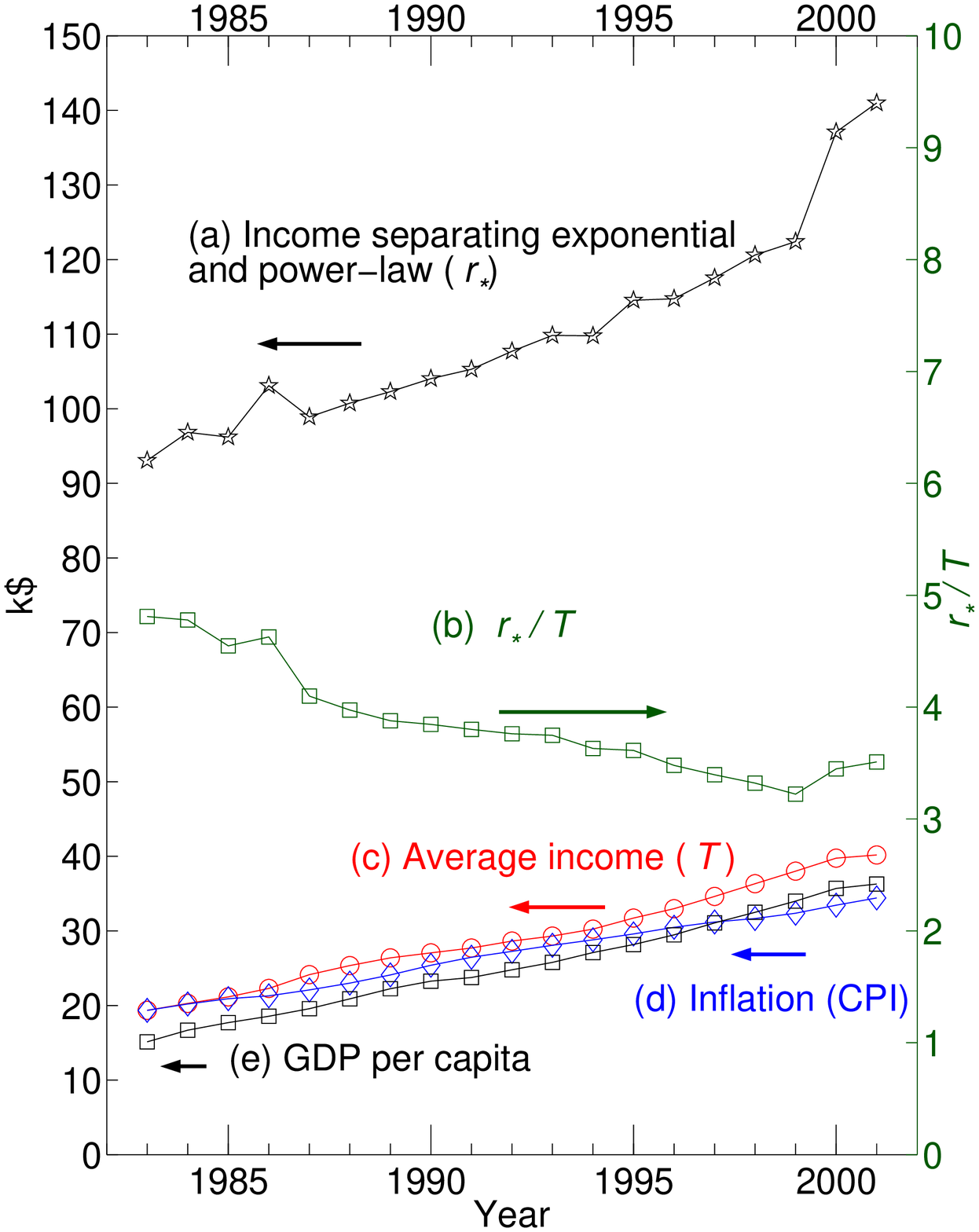}{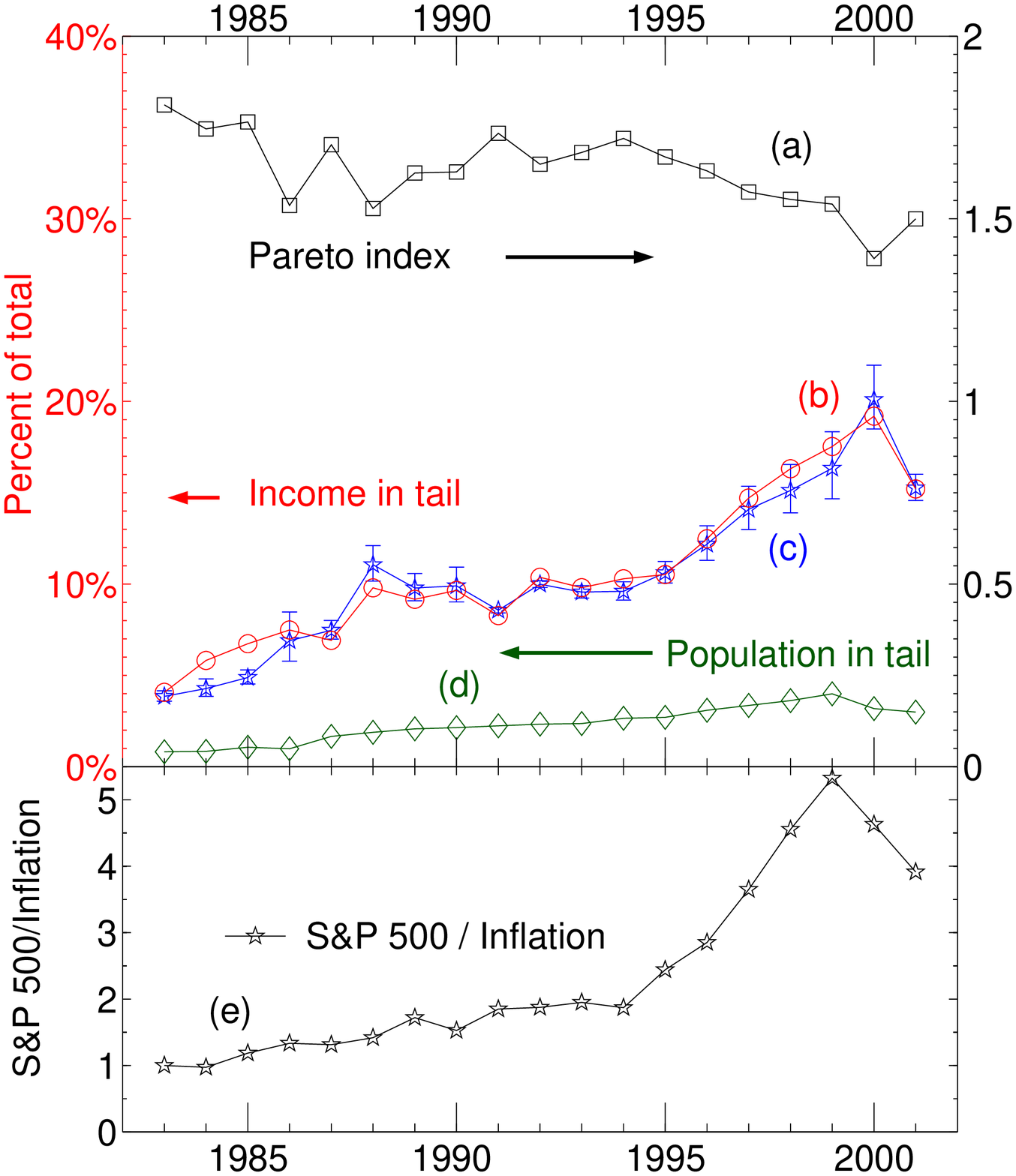}
\caption{
  Temporal evolution of various parameters characterizing income
  distribution.}
\label{fig:temperature}
\caption{
  (a) The Pareto index $\alpha$ of the power-law tail
  $C(r)\propto1/r^\alpha$.  (b) The excessive income in the Pareto
  tail, as a fraction $f$ of the total income in the system, obtained
  as the difference between the total income and the income in the
  exponential part of the distribution.  (c) The tail income fraction
  $f$, obtained by integrating the Pareto power law of the tail.  (d)
  The fraction of population belonging to the Pareto tail. (e) The
  stock-market index S\&P 500 divided by the inflation coefficient and
  normalized to 1 in 1983.}
\label{fig:index}
\end{figure}

Temporal evolution of the parameters $T$ and $r_*$ is shown in Fig.\ 
\ref{fig:temperature}.  We observe that the average income $T$ (in
nominal dollars) was increasing gradually, almost linearly in time,
and doubled in the last twenty years.  In Fig.\ \ref{fig:temperature},
we also show the inflation coefficient (the consumer price index CPI
from Ref.\ \cite{CPI}) compounded on the average income of 1983.  For
the twenty years, the inflation factor is about 1.7, thus most, if not
all, of the nominal increase in $T$ is inflation.  Also shown in Fig.\ 
\ref{fig:temperature} is the nominal gross domestic product (GDP) per
capita \cite{CPI}, which increases in time similarly to $T$ and CPI.
The ratio $r_*/T$ varies between 4.8 and 3.2 in Fig.\ 
\ref{fig:temperature}.

In Fig.\ \ref{fig:index}, we show how the parameters of the Pareto
tail $C(r)\propto 1/r^\alpha$ change in time.  Curve (a) shows that
the power-law index $\alpha$ varies between 1.8 and 1.4, so the power
law is not universal.  Because a power law decays with $r$ more slowly
than an exponential function, the upper tail contains more income than
we would expect for a thermal distribution, hence we call the tail
``superthermal'' \cite{superthermal}.  The total excessive income in
the upper tail can be determined in two ways: as the integral
$\int_{r_*}^{\infty}dr'r'P(r')$ of the power-law distribution, or as
the difference between the total income in the system and the income
in the exponential part.  Curves (c) and (b) in Fig.\ \ref{fig:index}
show the excessive income in the upper tail, as a fraction $f$ of the
total income in the system, determined by these two methods, which
agree with each other reasonably well.  We observe that $f$ increased
by the factor of 5 between 1983 and 2000, from 4\% to 20\%, but
decreased in 2001 after the crash of the US stock market.  For
comparison, curve (e) in Fig.\ \ref{fig:index} shows the stock market
index S\&P 500 divided by inflation.  It also increased by the factor
of 5.5 between 1983 and 1999, and then dropped after the stock market
crash.  We conclude that the swelling and shrinking of the upper
income tail is correlated with the rise and fall of the stock market.
Similar results were found for the upper income tail in Japan in Ref.\
\cite{Aoki}.  Curve (d) in Fig.\ \ref{fig:index} shows the fraction of
population in the upper tail.  It increased from 1\% in 1983 to 3\% in
1999, but then decreased after the stock market crash.  Notice,
however, that the stock market dynamics had a much weaker effect on
the average income $T$ of the lower, ``thermal'' part of income
distribution shown in Fig.\ \ref{fig:temperature}.

\begin{figure}
\includegraphics[width=0.55\linewidth,angle=-90]{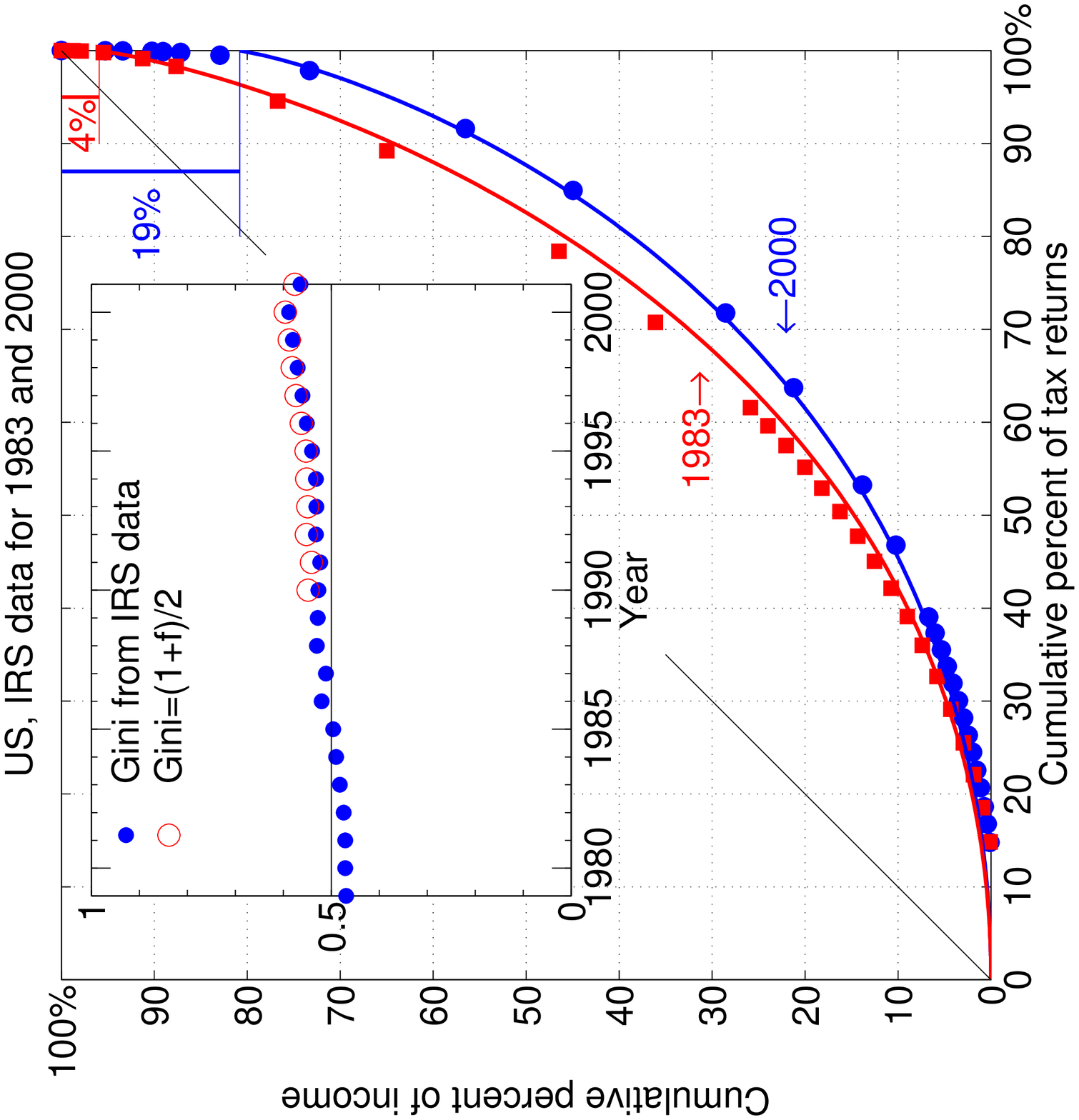}
\caption{
  Main panel: Lorenz plots for income distribution in 1983 and 2000.
  The data points are from the IRS \cite{Petska}, and the theoretical
  curves represent Eq.\ (\ref{eq:Lorenz}) with $f$ from Fig.\
  \ref{fig:index}.  Inset: The closed circles are the IRS data
  \cite{Petska} for the Gini coefficient $G$, and the open circles
  show the theoretical formula $G=(1+f)/2$.}
\label{fig:lorentz}
\end{figure}

For discussion of income inequality, the standard practice is to
construct the so-called Lorenz curve \cite{Kakwani}.  It is defined
parametrically in terms of the two coordinates $x(r)$ and $y(r)$
depending on the parameter $r$, which changes from 0 to $\infty$.  The
horizontal coordinate $x(r)=\int_{0}^{r}dr'P(r')$ is the fraction of
population with income below $r$.  The vertical coordinate
$y(r)=\int_0^rdr'r'P(r')/\int_{0}^{\infty}dr'r'P(r')$ is the total
income of this population, as a fraction of the total income in the
system.  Fig.\ \ref{fig:lorentz} shows the data points for the Lorenz
curves in 1983 and 2000, as computed by the IRS \cite{Petska}.  For a
purely exponential distribution of income $P(r)\propto\exp(-r/T)$, the
formula $y=x+(1-x)\ln(1-x)$ for the Lorenz curve was derived in Ref.\ 
\cite{Yakovenko-income}.  This formula describes income distribution
reasonably well in the first approximation \cite{Yakovenko-income},
but visible deviations exist.  These deviations can be corrected by
taking into account that the total income in the system is higher than
the income in the exponential part, because of the extra income in the
Pareto tail.  Correcting for this difference in the normalization of
$y$, we find a modified expression \cite{Yakovenko-survey} for the
Lorenz curve
\begin{equation}
  y=(1-f)[x+(1-x)\ln(1-x)]+f\Theta(x-1),
\label{eq:Lorenz}
\end{equation}
where $f$ is the fraction of the total income contained in the Pareto
tail, and $\Theta(x-1)$ is the step function equal to 0 for $x<1$ and
1 for $x\geq1$.  The Lorenz curve (\ref{eq:Lorenz}) experiences a
vertical jump of the height $f$ at $x=1$, which reflects the fact
that, although the fraction of population in the Pareto tail is very
small, their fraction $f$ of the total income is significant.  It does
not matter for Eq.\ (\ref{eq:Lorenz}) whether the extra income in the
upper tail is described by a power law or another slowly decreasing
function $P(r)$.  The Lorenz curves, calculated using Eq.\ 
(\ref{eq:Lorenz}) with the values of $f$ from Fig.\ \ref{fig:index},
fit the IRS data points very well in Fig.\ \ref{fig:lorentz}.

The deviation of the Lorenz curve from the diagonal in Fig.\
\ref{fig:lorentz} is a certain measure of income inequality.  Indeed,
if everybody had the same income, the Lorenz curve would be the
diagonal, because the fraction of income would be proportional to the
fraction of population.  The standard measure of income inequality is
the so-called Gini coefficient $0\leq G\leq1$, which is defined as the
area between the Lorenz curve and the diagonal, divided by the area of
the triangle beneath the diagonal \cite{Kakwani}.  It was calculated
in Ref.\ \cite{Yakovenko-income} that $G=1/2$ for a purely exponential
distribution.  Temporal evolution of the Gini coefficient, as
determined by the IRS \cite{Petska}, is shown in the inset of Fig.\
\ref{fig:lorentz}.  In the first approximation, $G$ is quite close to
the theoretically calculated value 1/2.  The agreement can be improved
by taking into account the Pareto tail, which gives $G=(1+f)/2$ for
Eq.\ (\ref{eq:Lorenz}).  The inset in Fig.\ \ref{fig:lorentz} shows
that this formula very well fits the IRS data for the 1990s with the
values of $f$ taken from Fig.\ \ref{fig:index}.  We observe that
income inequality was increasing for the last 20 years, because of
swelling of the Pareto tail, but started to decrease in 2001 after the
stock market crash.  The deviation of $G$ below 1/2 in the 1980s
cannot be captured by our formula.  The data points for the Lorenz
curve in 1983 lie slightly above the theoretical curve in Fig.\
\ref{fig:lorentz}, which accounts for $G<1/2$.

Thus far we discussed the distribution of individual income.  An
interesting related question is the distribution of family income
$P_2(r)$.  If both spouses are earners, and their incomes are
distributed exponentially as $P_1(r)\propto\exp(-r/T)$, then
\begin{equation}
  P_2(r)=\int_0^r dr' P_1(r')P_1(r-r')\propto r\exp(-r/T).
\label{eq:family}
\end{equation}
Eq.\ (\ref{eq:family}) is in a good agreement with the family income
distribution data from the US Census Bureau \cite{Yakovenko-income}.
In Eq.\ (\ref{eq:family}), we assumed that incomes of spouses are
uncorrelated.  This assumption was verified by comparison with the
data in Ref.\ \cite{Yakovenko-survey}.  The Gini coefficient for
family income distribution (\ref{eq:family}) was found to be
$G=3/8=37.5\%$ \cite{Yakovenko-income}, in agreement with the data.
Moreover, the calculated value 37.5\% is close to the average $G$ for
the developed capitalist countries of North America and Western
Europe, as determined by the World Bank \cite{Yakovenko-survey}.

On the basis of the analysis presented above, we propose a concept of
the \emph{equilibrium inequality} in a society, characterized by
$G=1/2$ for individual income and $G=3/8$ for family income.  It is a
consequence of the exponential Boltzmann-Gibbs distribution in thermal
equilibrium, which maximizes the entropy $S=\int dr\,P(r)\,\ln P(r)$
of a distribution $P(r)$ under the constraint of the conservation law
$\langle r\rangle=\int_0^\infty dr\,P(r)\,r=\rm const$.  Thus, any
deviation of income distribution from the exponential one, to either
less inequality or more inequality, reduces entropy and is not
favorable by the second law of thermodynamics.  Such deviations may be
possible only due to non-equilibrium effects.  The presented data show
that the great majority of the US population is in thermal
equilibrium.

Finally, we briefly discuss how the two-class structure of income
distribution can be rationalized on the basis of a kinetic approach,
which deals with temporal evolution of the probability distribution
$P(r,t)$.  Let us consider a diffusion model, where income $r$ changes
by $\Delta r$ over a period of time $\Delta t$.  Then, temporal
evolution of $P(r,t)$ is described by the Fokker-Planck equation
\cite{Kinetics}
\begin{equation}
   \frac{\partial P}{\partial t}=\frac{\partial}{\partial r}
   \left(AP + \frac{\partial}{\partial r}(BP)\right), \quad
   A=-{\langle\Delta r\rangle \over \Delta t}, \quad
   B={\langle(\Delta r)^2\rangle \over 2\Delta t}.
\label{eq:diffusion}
\end{equation}
For the lower part of the distribution, it is reasonable to assume
that $\Delta r$ is independent of $r$.  In this case, the coefficients
$A$ and $B$ are constants.  Then, the stationary solution
$\partial_tP=0$ of Eq.\ (\ref{eq:diffusion}) gives the exponential
distribution \cite{Yakovenko-money} $P(r)\propto\exp(-r/T)$ with
$T=B/A$.  Notice that a meaningful solution requires that $A>0$, i.e.\ 
$\langle\Delta r\rangle<0$ in Eq.\ (\ref{eq:diffusion}).  On the other
hand, for the upper tail of income distribution, it is reasonable to
expect that $\Delta r\propto r$ (the Gibrat law \cite{Gibrat}), so
$A=ar$ and $B=br^2$.  Then, the stationary solution $\partial_tP=0$ of
Eq.\ (\ref{eq:diffusion}) gives the power-law distribution
$P(r)\propto1/r^{\alpha+1}$ with $\alpha=1+a/b$.  The former process
is additive diffusion, where income changes by certain amounts,
whereas the latter process is multiplicative diffusion, where income
changes by certain percentages.  The lower class income comes from
wages and salaries, so the additive process is appropriate, whereas
the upper class income comes from investments, capital gains, etc.,\ 
where the multiplicative process is applicable.  Ref.\ \cite{Aoki}
quantitatively studied income kinetics using tax data for the upper
class in Japan and found that it is indeed governed by a
multiplicative process.  The data on income mobility in the USA are
not readily available publicly, but are accessible to the Statistics
of Income Research Division of the IRS.  Such data would allow to
verify the conjectures about income kinetics.

The exponential probability distribution $P(r)\propto\exp(-r/T)$ is a
monotonous function of $r$ with the most probable income $r=0$.  The
probability densities shown in Fig.\ \ref{fig:LogLin} agree reasonably
well with this simple exponential law.  However, a number of other
studies found a nonmonotonous $P(r)$ with a maximum at $r\neq0$ and
$P(0)=0$.  These data were fitted by the log-normal
\cite{Souma,Gallegati,Australia} or the gamma distribution
\cite{Mimkes,West,Ferrero}.  The origin of the discrepancy in the
low-income data between our work and other papers is not completely
clear at this moment.  The following factors may possibly play a role.
First, one should be careful to distinguish between personal income
and group income, such as family and household income.  As Eq.\ 
(\ref{eq:family}) shows, the later is given by the gamma distribution
even when the personal income distribution is exponential.  Very often
statistical data are given for households and mix individual and group
income distributions (see more discussion in Ref.\ 
\cite{Yakovenko-income}).  Second, the data from tax agencies and
census bureaus may differ.  The former data are obtained from tax
declarations of all taxable population, whereas the later data from
questionnaire surveys of a limited sample of population.  These two
methodologies may produce different results, particularly for low
incomes.  Third, it is necessary to distinguish between distributions
of money \cite{Yakovenko-money,Ferrero,Chakrabarti}, wealth
\cite{West,wealth}, and income.  They are, presumably, closely
related, but may be different in some respects.  Fourth, the low-income
probability density may be different in the USA and in other countries
because of different social security policies.  All these questions
require careful investigation in future work.  We can only say that
the data sets analyzed in this paper and our previous papers are well
described by a simple exponential function for the whole lower class.
This does not exclude a possibility that other functions can also fit
the data \cite{Dragulescu}.  However, the exponential law has only one
fitting parameter $T$, whereas log-normal, gamma, and other
distributions have two or more fitting parameters, so they are less
parsimonious.

\vspace{-1\baselineskip}

\end{document}